\documentclass{iopjournal}

\usepackage{amsfonts}
\usepackage{amsmath}
\usepackage{amssymb}
\usepackage{bm}

\begin{document}

\articletype{Paper} 

\title{Optical signatures of Rashba--Dresselhaus 
coupling across Franz--Keldysh and Landau regimes 
in two-dimensional semiconductors}

\author{Bahruz Suleymanli$^1$\orcid{0000-0001-5700-0464}, 
Samir Tariverdiev$^2$\orcid{0009-0008-6675-0150} 
and B. Tanatar$^{3,*}$\orcid{0000-0002-5246-0119}}

\affil{$^1$Department of Physics, Yıldız Technical 
University, 34220 Esenler, Istanbul, Türkiye}

\affil{$^2$Sivas Cumhuriyet University Nanophotonics 
Application and Research Center-CÜNAM, 58140 Sivas, 
Türkiye}

\affil{$^3$Department of Physics, Bilkent University, 
06800 Ankara, Türkiye}

\affil{$^*$Author to whom any correspondence should be addressed.}

\email{bahruz.suleymanli@gmail.com, samirtariverdiyev@cumhuriyet.edu.tr 
and tanatar@fen.bilkent.edu.tr}


\begin{abstract}
We develop a unified theory of the interband optical 
response of a two-dimensional semiconductor with 
simultaneous Rashba and Dresselhaus spin-orbit 
couplings in electric and magnetic fields. In 
crossed fields, the in-plane electric field 
generates an effective spin-orbit drift 
field that modifies the Landau spectrum and 
spinor states, while at zero magnetic field 
the problem maps onto Landau--Zener spin 
dynamics.
At zero external fields, the total joint 
density of states above the band gap 
remains unchanged, while characteristic 
subgap structures appear. Pure Rashba 
coupling produces an inverse square root 
singularity, whereas a finite Dresselhaus 
term gives a step and a logarithmic peak, 
allowing the two couplings to be 
distinguished optically. At zero magnetic 
field, the Franz--Keldysh response shows 
a crossover from diabatic to adiabatic 
spin dynamics as the electric field is 
reduced. In the weak-field Rashba regime, 
the field-free singularity becomes a 
finite peak whose height scales as 
$\mathcal E^{-1/3}$ and whose position 
approaches $E_g-E_{\rm SO}$, providing a 
direct measure of the Rashba energy. In a 
perpendicular magnetic field, spin-orbit 
coupling shifts and splits the Landau 
absorption lines and produces anticrossings 
without changing their total weights. The 
Rashba splitting changes from a $\sqrt{B}$ 
dependence at weak fields to a linear 
dependence at strong fields. Finally, in 
crossed electric and magnetic fields, the 
electric field broadens the Landau lines 
and drives a continuous crossover toward 
the Franz--Keldysh regime, while spin-orbit 
coupling produces opposite shifts and 
different widths for the two members of 
each doublet.
\end{abstract}

\section{\label{sec:intro}Introduction}

Spin-orbit coupling connects the motion of charge 
carriers with their spin and has therefore become 
a central mechanism for spin control in low-dimensional 
semiconductor systems~\cite{Zutic2004,Awschalom2007,
Manchon2015,Avsar2020,Bihlmayer2022,Zollner2025}. 
In two-dimensional semiconductor heterostructures, 
the dominant linear-in-momentum spin-orbit interactions are commonly 
the Rashba and Dresselhaus terms. The Rashba interaction 
originates from an asymmetry of the confining structure 
or potential, while the Dresselhaus interaction 
originates from the lack of inversion symmetry of the 
crystal lattice. Both interactions can coexist and their 
strengths can be measured and controlled experimentally
~\cite{Nitta1997,Meier2007,Nishimura2021,Chander2026}. 
The relative strength of these two couplings determines 
the spin structure of the system and leads to a special 
symmetry when the Rashba and Dresselhaus coupling 
strengths become equal, where the persistent 
spin helix appears~\cite{Schliemann2003_1,Bernevig2006,Koralek2009,
Tkach2022}. The observation of giant Rashba splittings 
in polar semiconductors has also shown that the 
spin-orbit energy can become comparable to the energy scales 
associated with experimentally accessible electric and 
magnetic fields~\cite{Ishizaka2011,Ketterl2021,Lee2022}. 
In this regime, the optical response can be strongly 
affected by all these mechanisms. An electric field 
modifies the interband absorption through the 
Franz--Keldysh effect, while a perpendicular magnetic 
field produces Landau quantization and a discrete 
magnetoabsorption spectrum~\cite{Miller1986,Citrin2000,
Sercel2023,Yudin2016}. It is therefore important to 
understand the optical response when Rashba and 
Dresselhaus couplings act together in the presence of 
both in-plane electric and perpendicular magnetic fields.

This combined regime is not only of theoretical interest, 
but can be accessed experimentally in a number of 
low-dimensional semiconductor platforms. Among these systems, 
III--V quantum wells are particularly important because 
their electronic, spin, and optical properties can be 
engineered within the same heterostructure. InAs and 
InSbAs wells combine high mobility two-dimensional 
electron gases with strong spin-orbit coupling and 
external field control~\cite{Lee2021,Metti2022,
Stier2023}, while GaAs/AlGaAs and InGaAs structures 
allow the confinement, optical transitions, and 
spin-orbit interaction to be tuned through the layer and 
interface design~\cite{Studer2009,Alekseev2021}.
Their direct band gaps and mature growth technology 
also make these materials important for semiconductor 
lasers and integrated photonic devices~\cite{Mobini2022,
Baboux2023}.
Halide perovskites provide 
another important platform because strong spin-orbit 
effects can coexist with optical activity and electric 
field control. In these materials, electrically switchable 
spin structures have been demonstrated at room temperature, 
while magneto-optical experiments have shown that magnetic 
fields can strongly modify the optical response~\cite{
Zhang2022,Mutzafi2022}. More recently, 
Rashba--Dresselhaus type spin physics has also been 
realized in exciton polaritons, which are mixed light 
and matter states, allowing spin-dependent optical 
transport and a spin Hall effect at room temperature
~\cite{Liang2024}. These developments are directly 
relevant for spin transistors, spin-orbit torque 
memories, optical spin control, and integrated opto 
spintronic devices, where electric fields, magnetic 
fields, spin-orbit coupling, and light must act 
within the same low-dimensional system~\cite{Ahn2020,
Dieny2020,Yang2022,Gish2024}.

Existing theoretical studies show that the combined 
Rashba and Dresselhaus spectrum in a perpendicular 
magnetic field, including the effect of an in-plane 
electric field, has mainly been studied at the level 
of Landau energies and spin structure~\cite{Lipparini2006,
Papp2010, Suleymanli2022, Candido2023}. 
However, analytical treatments of this problem have often 
relied on approximate or perturbative solutions, especially 
when the in-plane electric field is included, and the role 
of this field as a spin-orbit drift field together with its 
effect on the crossings, anticrossings, and mixing of the 
Landau states has received much less attention.
Building on this, optical studies have considered either 
the zero field Rashba--Dresselhaus response~\cite{Li2013} 
or magnetoabsorption in a perpendicular field, mostly for 
Rashba coupling alone~\cite{Sercel2023,Yang2008}. Optical 
absorption in crossed electric and magnetic fields has also 
been studied, but generally without spin-orbit coupling
~\cite{Spector1964,Vrehen1966,Monozon2019}, while calculations 
including both spin-orbit couplings and both fields have been 
limited to confined structures such as quantum rings
~\cite{Seyed2018}. Therefore, the interband optical response 
of an extended two-dimensional system with simultaneous 
Rashba and Dresselhaus couplings, an in-plane electric field, 
and a perpendicular magnetic field remains largely unexplored. 
In particular, a common description of the Franz--Keldysh 
response, the magnetoabsorption spectrum, and the crossover 
between these two regimes is still missing.

In this work, we develop a unified description of the 
interband optical response of a two-dimensional 
semiconductor with simultaneous Rashba and Dresselhaus 
couplings under an 
in-plane electric field and a perpendicular magnetic field. 
In Sec.~\ref{sec:2}, we introduce the model and formulate 
the optical absorption in terms of the joint density of 
states. The crossed field spectrum and spinor states are 
obtained by numerically converged exact diagonalization in 
Sec.~\ref{sec:3a}, where the electric field induced 
spin-orbit drift term and the 
resulting level shifts and anticrossings are analyzed.
In Sec.~\ref{sec:3b}, the zero magnetic field spin dynamics 
is mapped onto the Landau--Zener problem, including the 
persistent spin helix limit. The field-free optical 
response is studied in Sec.~\ref{sec:3c}, followed by 
the Franz--Keldysh electroabsorption in Sec.~\ref{sec:3d} 
and the magnetoabsorption spectrum in Sec.~\ref{sec:3e}. 
Finally, Sec.~\ref{sec:3f} treats the full cross-field 
problem. The main conclusions are summarized in 
Sec.~\ref{sec:Conclusions}, where we also discuss 
representative material parameters and the experimental 
accessibility of the predicted effects.

\section{\label{sec:2}Model of spin–orbit coupled 
electroabsorption}

\begin{figure*}
    \centering
    \includegraphics[width=1\textwidth]{
    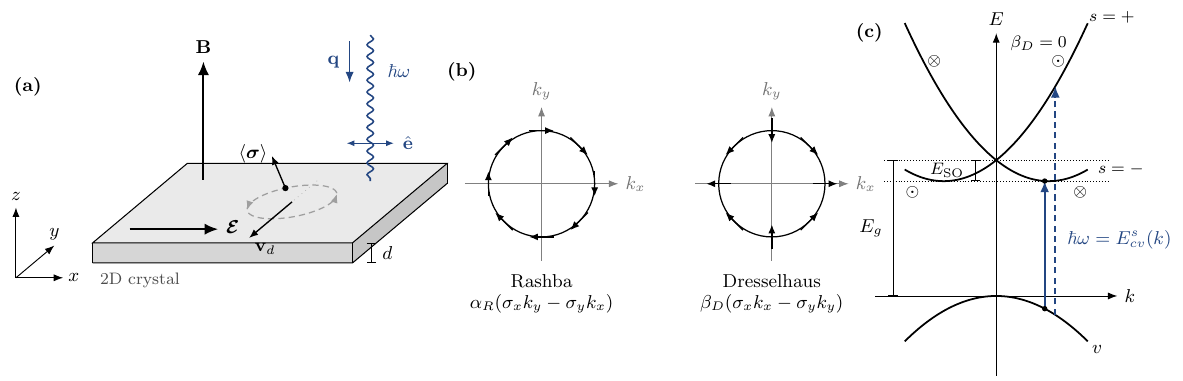}
    \caption{Schematic representation of the 
    model. (a) A 2D crystal of effective 
    thickness $d$ subjected to a perpendicular 
    magnetic field $\mathbf B\parallel \hat z$ 
    and an in-plane electric field 
    $\bm{\mathcal E}\parallel \hat x$, together 
    with normally incident monochromatic light 
    of angular frequency $\omega$, wave vector 
    $\mathbf q\parallel \hat z$, and in-plane 
    polarization $\hat{\mathbf e}$. The dashed 
    orbit shows the cyclotron motion, $\mathbf 
    v_d=c\,\bm{\mathcal E}\times\mathbf B/B^2$ 
    denotes the Hall drift velocity, and 
    $\langle\boldsymbol{\sigma}\rangle$ indicates 
    the spin orientation. (b) In-plane spin 
    orientations for Rashba and Dresselhaus 
    couplings on constant-energy contours. 
    (c) Direct interband optical transitions with 
    $\hbar\omega=E_{cv}^{s}(k)$ for the pure Rashba 
    limit $\beta_D=0$, where $s=\pm$ labels the two 
    spin branches, $\odot$ and $\otimes$ indicate 
    in-plane spins pointing out of and into the 
    plane of the figure, $E_g$ is the band gap, 
    and $E_{\rm SO}=m^*\alpha_R^2/2$ is the Rashba 
    spin--orbit energy.}
    \label{fig:model}
\end{figure*}

We consider a 2D crystal in the $xy$ plane in the 
presence of Rashba and Dresselhaus spin-orbit 
coupling (SOC), subjected to a magnetic field 
$\mathbf{B}=(0,0,B)$ normal to the plane and 
an electric field $\bm{\mathcal E}=(\mathcal 
E,0,0)$ applied along the $x$ direction. 
The crystal is exposed to monochromatic light 
of angular frequency $\omega$, whose photons 
induce direct interband transitions from the 
valence band to the conduction band, as 
schematically shown in Fig.~\ref{fig:model}.

Denoting the complex dielectric function of the 
crystal by $\epsilon=\epsilon_1+i\epsilon_2$ and 
the electric-field amplitude of the light wave 
inside the layer by $E_\perp$, 
the time-averaged absorbed power per unit area 
of a 2D layer with effective thickness $d$ can 
be written on the one hand as
$w=\frac{1}{2}\omega\epsilon_2\epsilon_0 d 
E_\perp^2$
and on the other hand as
$w=\hbar\omega w_{cv}$,
where $\epsilon_0$ is the vacuum permittivity 
and $w_{cv}$ is the interband transition rate 
per unit area. Using Fermi's golden rule for a 
direct interband dipole transition from the 
valence band $v$ to the conduction band $c$, we 
can write the imaginary part of the dielectric 
function as
$\epsilon_2(\omega)=
\pi e^2
\left|\hat{\mathbf e}\cdot\mathbf p_{cv}\right|^2
J_{cv}(\hbar\omega)/
\left(\epsilon_0m_0^2\omega^2d\right)$,
where $\hat{\mathbf e}$ is the polarization vector, 
$\mathbf p_{cv}$ is the interband momentum matrix 
element, $m_0$ is the free electron mass, and 
$J_{cv}$ is the joint density of states per unit 
area and per unit energy. Knowing 
$\epsilon_2(\omega)$, the absorption coefficient  
can be obtained as 
$\omega\epsilon_2(\omega)/(n_0c)$, where $n_0$ is 
the background refractive index. The absorbance 
$\mathcal A(\omega)$ is then given by
\begin{equation}
\mathcal A(\omega)=
\frac{\pi e^2}{\epsilon_0 n_0 c m_0^2\omega}
\left|\hat{\mathbf e}\cdot\mathbf p_{cv}\right|^2
J_{cv}(\hbar\omega).
\label{eq:absorbance}
\end{equation}
We employ the constant-matrix-element approximation 
and neglect the momentum dependence of $\mathbf p_{cv}$. 
As can be seen from Eq.~\eqref{eq:absorbance}, the 
entire dependence of the optical absorption on the 
SOCs and external fields is therefore contained in 
$J_{cv}$. For noninteracting electrons and holes, 
the joint density of states can be written as
\begin{equation}
J_{cv}(\hbar\omega)=
\sum_{\nu}
\left|\Psi_{\nu}(\mathbf r=0)\right|^2
\delta\left(\hbar\omega-E_g-E_{\nu}\right),
\label{eq:jcv}
\end{equation}
where $E_g$ is the band gap, while $E_{\nu}$ and 
$\Psi_{\nu}$ are the eigenvalues and normalized 
eigenfunctions of the electron--hole relative 
motion, with $\mathbf r=\mathbf r_e-\mathbf r_h$ 
denoting the relative coordinate. To determine 
these quantities, we describe the relative motion 
within the effective-mass approximation by the 
Hamiltonian
\begin{equation}
H=\frac{\Pi_x^2+\Pi_y^2}{2m^{*}}
-e\mathcal{E}x
+\frac{g}{2}\mu_BB\sigma_z
+\alpha_R\left[\sigma_x\Pi_y-\sigma_y\Pi_x\right]
+\beta_D\left[\sigma_x\Pi_x-\sigma_y\Pi_y\right],
\label{eq:H}
\end{equation}
where $\mathbf{\Pi}=-i\hbar\mathbf{\nabla}-e\mathbf A$,
$1/m^*=1/m^*_e+1/m^*_h$
defines the reduced effective mass of the electron--hole 
pair, $g$ is the effective Land\'e factor, $\mu_B$ 
is the Bohr magneton, $\sigma_i$ are the Pauli matrices, 
and $\alpha_R$ and $\beta_D$ are the effective Rashba 
and Dresselhaus coupling constants, respectively. We 
take $e=-|e|$ throughout.
In Eq.~\eqref{eq:H} the Zeeman and spin-orbit terms 
act on the relative motion with effective constants. 
This description reduces to the standard Landau-level 
and Franz--Keldysh descriptions of interband absorption 
when the SOC is switched off, and it is exact in the 
absence of fields when only one of the two bands carries 
a linear spin splitting.

For the considered field configuration we use the 
Landau gauge
$\mathbf A=B(0,x,0)$. In this gauge, translational 
invariance along
$y$ is preserved and $k_y$ remains a good quantum 
number. The general state index $\nu$ in 
Eq.~\eqref{eq:jcv} can therefore be represented by 
the level index $n$ and the conserved wave vector 
$k_y$. For the Hamiltonian given in Eq.~\eqref{eq:H}, 
the stationary Schr\"odinger equation is written as
$H\Psi_{nk_y}(x,y)=E_{nk_y}\Psi_{nk_y}(x,y)$,
and we write the spinor wave function as
$\Psi_{nk_y}(x,y)=L_y^{-1/2}e^{ik_yy}\psi_{nk_y}(x)$, 
where $L_y$ is the length of the crystal along $y$.

Introducing the dimensionless coordinate $z=x/a_B$, 
where
$a_B=\sqrt{\hbar/\left(m^{*}\omega_B\right)}$
is the magnetic length and
$\omega_B=|e|B/m^{*}$
is the cyclotron frequency, and measuring energies 
in units of $\hbar\omega_B/2$, the Schr\"odinger 
equation becomes 
\begin{equation}
\Bigg\{
\frac{d^2}{dz^2}
-(z-z_0)^2
-2\tilde{\mathcal E}z
-\sigma_z\mu_Z
-\alpha\left[
\sigma_x(z-z_0)
+i\sigma_y\frac{d}{dz}
\right]
+i\beta\left[
\sigma_x\frac{d}{dz}
-i\sigma_y(z-z_0)
\right]
+\tilde E_{nk_y}
\Bigg\}\psi_{nk_y}(z)=0,
\label{eq:dimless}
\end{equation}
where the dimensionless couplings, fields and energies are
\begin{equation}
\alpha=\sqrt{\frac{4m^{*}\alpha_R^2}{\hbar\omega_B}},
\quad
\beta=\sqrt{\frac{4m^{*}\beta_D^2}{\hbar\omega_B}},
\quad
\tilde{\mathcal E}=\frac{|e|\mathcal{E}a_B}{\hbar\omega_B},
\quad
\mu_Z=\frac{g\mu_BB}{\hbar\omega_B},
\quad
\tilde E_{nk_y}=\frac{E_{nk_y}}{\hbar\omega_B/2},
\quad
z_0=-k_ya_B .
\label{eq:params}
\end{equation}
Here $z_0$ is the dimensionless coordinate of the center 
of the cyclotron orbit, and $\tilde{\mathcal E}
=m^*v_da_B/\hbar$ is the drift momentum $m^*v_d$, 
with $v_d=\mathcal E/B$ the Hall drift velocity, in units of 
$\hbar/a_B$. 
To write Eq.~\eqref{eq:dimless} in Hermite form for the spinor 
components
$\psi_{nk_y}=(\psi_{n\uparrow},\psi_{n\downarrow})^T$, 
we introduce
$\eta=z-z_0+\tilde{\mathcal E}$, measured from the displaced orbit 
center, and factor out the Gaussian by writing
$\psi_{n\uparrow,n\downarrow}(z)=e^{-\eta^2/2}
\varphi_{n\uparrow,n\downarrow}(\eta)$.
We then obtain
\begin{eqnarray}
\left\{
\frac{d^2}{d\eta^2}
-2\eta\frac{d}{d\eta}
+\mathbb E_n-\mu_Z
\right\}
\varphi_{n\uparrow}(\eta)
-\left\{
\alpha\left(
\frac{d}{d\eta}-\tilde{\mathcal E}
\right)
-i\beta\left(
\frac{d}{d\eta}-2\eta+\tilde{\mathcal E}
\right)
\right\}
\varphi_{n\downarrow}(\eta)=0,
\nonumber\\
\left\{
\frac{d^2}{d\eta^2}
-2\eta\frac{d}{d\eta}
+\mathbb E_n+\mu_Z
\right\}
\varphi_{n\downarrow}(\eta)
+\left\{
\alpha\left(
\frac{d}{d\eta}-2\eta+\tilde{\mathcal E}
\right)
+i\beta\left(
\frac{d}{d\eta}-\tilde{\mathcal E}
\right)
\right\}
\varphi_{n\uparrow}(\eta)=0.
\label{eq:hermite}
\end{eqnarray}
where the reduced energy is
$\mathbb E_n=
\tilde E_{nk_y}
+\tilde{\mathcal E}
\left(
\tilde{\mathcal E}-2z_0
\right)
-1.$
Eqs.~\eqref{eq:hermite} no longer contain $z_0$, so that 
the reduced energy $\mathbb E_n$ is independent of $k_y$ 
and the entire $k_y$ dependence of the spectrum is 
contained in the definition of $\mathbb E_n$, i.e.,
\begin{equation}
E_{nk_y}=\frac{\hbar\omega_B}{2}\left(\mathbb E_n+1\right)
-\frac{m^*v_d^2}{2}+|e|\mathcal E x_c,  
\label{eq:Enky}
\end{equation}
where $x_c=-k_ya_B^2$ is the center of the cyclotron 
orbit in the absence of the electric field. Every 
level is rigidly shifted by the electric potential 
at the orbit center and lowered by the drift kinetic 
energy, exactly as in the spin-degenerate 
crossed-field problem.

Eqs.~\eqref{eq:hermite} are solved by expanding the spinor 
components in normalized Hermite functions
$h_j(\eta)=H_j(\eta)/\left(\sqrt{\pi}\,2^j j!\right)^{1/2}$,
\begin{equation}
\varphi_{n\uparrow}(\eta)
=
\sum_{j=0}^{\infty}a_j^{(n)}h_j(\eta),
\quad
\varphi_{n\downarrow}(\eta)
=
\sum_{j=0}^{\infty}b_j^{(n)}h_j(\eta),
\label{eq:expansion}
\end{equation}
so that the full wave function is
$\psi_{nk_y}(x)=a_B^{-1/2}\sum_j(a_j^{(n)},b_j^{(n)})^T\phi_j(\eta)$, 
with the oscillator functions
$\phi_j=e^{-\eta^2/2}h_j$
and the normalization
$\sum_j\left(|a_j^{(n)}|^2+|b_j^{(n)}|^2\right)=1$.
Using
$H_j''-2\eta H_j'+2jH_j=0$,
$H_j'=2jH_{j-1}$
and
$H_j'-2\eta H_j=-H_{j+1}$,
the projection of Eqs.~\eqref{eq:hermite} onto $h_j$ yields the
three-term recursion
\begin{align}
\mathbb E_n a_j^{(n)}
={}&
(2j+\mu_Z)a_j^{(n)}
-(\alpha+i\beta)\tilde{\mathcal E}\,b_j^{(n)}
+\alpha\sqrt{2(j+1)}\,b_{j+1}^{(n)}
+i\beta\sqrt{2j}\,b_{j-1}^{(n)},
\nonumber\\
\mathbb E_n b_j^{(n)}
={}&
(2j-\mu_Z)b_j^{(n)}
-(\alpha-i\beta)\tilde{\mathcal E}\,a_j^{(n)}
-i\beta\sqrt{2(j+1)}\,a_{j+1}^{(n)}
+\alpha\sqrt{2j}\,a_{j-1}^{(n)}.
\label{eq:recursion}
\end{align}
Thus, the field and SOC dependence of the electroabsorption is
determined by the eigenvalues $\mathbb E_n$ and coefficients 
$a_j^{(n)}$ and $b_j^{(n)}$ obtained from Eqs.~\eqref{eq:recursion},
which determine the energies $E_{nk_y}$ and wave functions
$\Psi_{nk_y}$ entering Eq.~\eqref{eq:jcv}. Since the couplings
in Eqs.~\eqref{eq:recursion} connect only neighboring Hermite
indices, the coefficient matrix is a Hermitian block-tridiagonal
matrix in which $\mathbb E_n$ appears only on the diagonal, 
so that the reduced energies and the expansion coefficients 
follow from a single diagonalization of the truncated matrix
rather than from the roots of a secular determinant. Because
the truncation error decreases exponentially with the number 
$N$ of retained Hermite functions, the lowest levels are 
obtained to machine precision already for $N$ of the order
of one hundred, and the closed-form limits of the problem 
are reproduced to the same accuracy. In this sense the solution
is exact.


\section{\label{sec:3}Results}

We now present the results that follow from 
Eqs.~\eqref{eq:recursion} and \eqref{eq:jcv}. The 
spin-orbit coupled energy spectrum in crossed fields is 
discussed first, together with its limiting cases, 
because it fixes the positions and the shapes of all 
absorption lines. The case of a vanishing magnetic 
field, in which the Hermite expansion is no longer 
applicable, is treated separately. The joint density 
of states is then obtained for different Rashba and 
Dresselhaus couplings in the four field configurations, 
namely without fields, in a pure electric field, in a 
pure magnetic field, and in crossed fields.

\subsection{\label{sec:3a}Spin-orbit coupled energy 
spectrum in crossed fields}

It is useful first to emphasize that 
Eqs.~\eqref{eq:recursion} can be written as the 
eigenvalue problem
$\mathbb H\varphi_n=\mathbb E_n\varphi_n$, where
$\mathbb H=
2b^\dagger b
+\mu_Z\sigma_z
+\sqrt{2}\alpha
\left(
b\sigma_+
+b^\dagger\sigma_-
\right)
+i\sqrt{2}\beta
\left(
b^\dagger\sigma_+
-b\sigma_-
\right)
-\tilde{\mathcal E}
\left(
\alpha\sigma_x-\beta\sigma_y
\right)$,
$b$ and $b^\dagger$ are the ladder operators of the 
oscillator functions $\phi_j$, with
$b\phi_j=\sqrt{j}\phi_{j-1}$ and
$b^\dagger\phi_j=\sqrt{j+1}\phi_{j+1}$, while
$\sigma_\pm=(\sigma_x\pm i\sigma_y)/2$ are the spin 
raising and lowering operators. As can be seen from 
the expression for $\mathbb H$, the first four terms 
describe the Landau problem with Rashba and Dresselhaus 
couplings, in which the Rashba term connects the states 
$|j,\uparrow\rangle$ and $|j+1,\downarrow\rangle$, while 
the Dresselhaus term connects $|j,\uparrow\rangle$ and 
$|j-1,\downarrow\rangle$. The electric field enters only 
through the last term. This term has the form of a 
Zeeman coupling to an effective magnetic field lying 
in the plane and does not change the oscillator index. 
Restoring dimensions through
$(\hbar\omega_B/2)\tilde{\mathcal E}\alpha =m^*v_d
\alpha_R$,
this term can be written as
$-m^*v_d(\alpha_R\sigma_x-\beta_D\sigma_y)$.
From this expression, it is clearly seen that the orbit 
center of an electron in crossed fields drifts with 
the Hall velocity $v_d$ along $-y$, so that the electron 
carries the drift momentum $m^*v_d$, and the spin orbit 
couplings convert this momentum into an effective in 
plane field, in the same way as a free electron with 
momentum $\hbar\mathbf k$ experiences the Rashba field 
$\hbar\alpha_R(\mathbf k\times\hat{\mathbf z})$. Apart 
from the classical drift terms of Eq.~\eqref{eq:Enky}, 
which are the same for all levels, this drift field 
is the only way in which the in-plane electric field 
acts on the spin of the electron. Thus, for an infinite 
crystal the nontrivial electric field dependence of the 
spin-orbit-coupled Landau spectrum is generated entirely 
by the drift field, which mixes neighboring Landau 
states, modifies the level structure, and displaces 
and deforms the spinor densities $|\chi_n(\eta)|^2=
e^{-\eta^2}\left(|\varphi_{n\uparrow}(\eta)|^2+
|\varphi_{n\downarrow}(\eta)|^2\right)$, with 
$\int |\chi_n(\eta)|^2 d\eta=1$, which, as shown in 
Sec.~\ref{sec:3f}, determine directly the 
electroabsorption line shapes.

We next consider several limiting cases in which 
$\mathbb H$ can be diagonalized in closed form. 
We first examine the case of pure Rashba coupling 
without an electric field, $\beta=\tilde{\mathcal E}
=0$. In this limit, the recursion decouples into 
$2\times2$ blocks formed by $a_j$ and $b_{j+1}$, 
together with the isolated coefficient $b_0$, and 
we obtain
\begin{eqnarray}
E_j^{\pm}
&=&
\hbar\omega_B(j+1)\pm
\sqrt{
\left(
\frac{\hbar\omega_B-g\mu_BB}{2}
\right)^2
+
4E_{\rm SO}\hbar\omega_B(j+1)
},\ \ \
\label{eq:ByRas}
\end{eqnarray}
together with the isolated level
$E_{\rm iso}=(\hbar\omega_B-g\mu_BB)/2$, where 
$E_{\rm SO}=m^*\alpha_R^2/2$ is the Rashba energy 
introduced in Fig.~\ref{fig:model}(c). This is the 
Bychkov--Rashba spectrum \cite{Bychkov1984}.
For pure Dresselhaus coupling, $\alpha=
\tilde{\mathcal E}=0$, the blocks are instead formed 
by $a_{j+1}$ and $b_j$, while $a_0$ remains isolated. 
The spectrum is obtained from the Rashba result by 
the replacements $\alpha\rightarrow\beta$ and 
$\mu_Z\rightarrow-\mu_Z$, with the isolated level 
$E_{\rm iso}=(\hbar\omega_B+g\mu_BB)/2$. The Rashba 
and Dresselhaus spectra are therefore equivalent 
only for $\mu_Z=0$. In the absence of SOC, the 
reduced spectrum becomes $\mathbb E_j=2j\pm\mu_Z$ 
for any electric field. Using Eq.~\eqref{eq:Enky}, 
the corresponding crossed-field Landau levels are
\begin{equation}
E_{nk_y}
=
\hbar\omega_B\left(j+\frac{1}{2}\right)
\pm\frac{g\mu_BB}{2}
-\frac{m^*v_d^2}{2}
-|e|\mathcal E k_ya_B^2,
\label{eq:noSOC}
\end{equation}
which is the result for the spin-degenerate crossed 
field problem.

It is important to note here that, for a weak electric 
field the drift field $-m^*v_d(\alpha_R\sigma_x
-\beta_D\sigma_y)$ can be treated as a perturbation. 
The field-free part of $\mathbb H$ changes the 
oscillator index by one whenever it flips the spin, 
so that each field-free eigenstate contains spin up 
components of one parity of $j$ and spin down 
components of the opposite parity, whereas 
$\sigma_x$ and $\sigma_y$ connect components with 
the same $j$. The drift field therefore has no 
diagonal matrix elements in the field-free eigenbasis, 
and every nondegenerate level shifts quadratically, 
$\delta\mathbb E_n\sim\tilde{\mathcal E}^2(\alpha^2
+\beta^2)$, i.e., $\delta E_n\sim(m^*v_d)^2(\alpha_R^2
+\beta_D^2)/\hbar\omega_B$ in physical units, with 
coefficients that are of opposite sign for the two 
members of a spin-orbit doublet in all cases studied 
below. Levels that are degenerate or nearly 
degenerate at $\tilde{\mathcal E}=0$ are instead 
split in first order by the matrix element of the 
drift field between them. This is a spin-orbit 
induced quadratic Stark effect of the Landau levels. 
It is absent without spin-orbit coupling, where 
Eq.~\eqref{eq:noSOC} contains no level dependent 
field term.

\begin{figure*}
    \centering
    \includegraphics[width=1\textwidth]
    {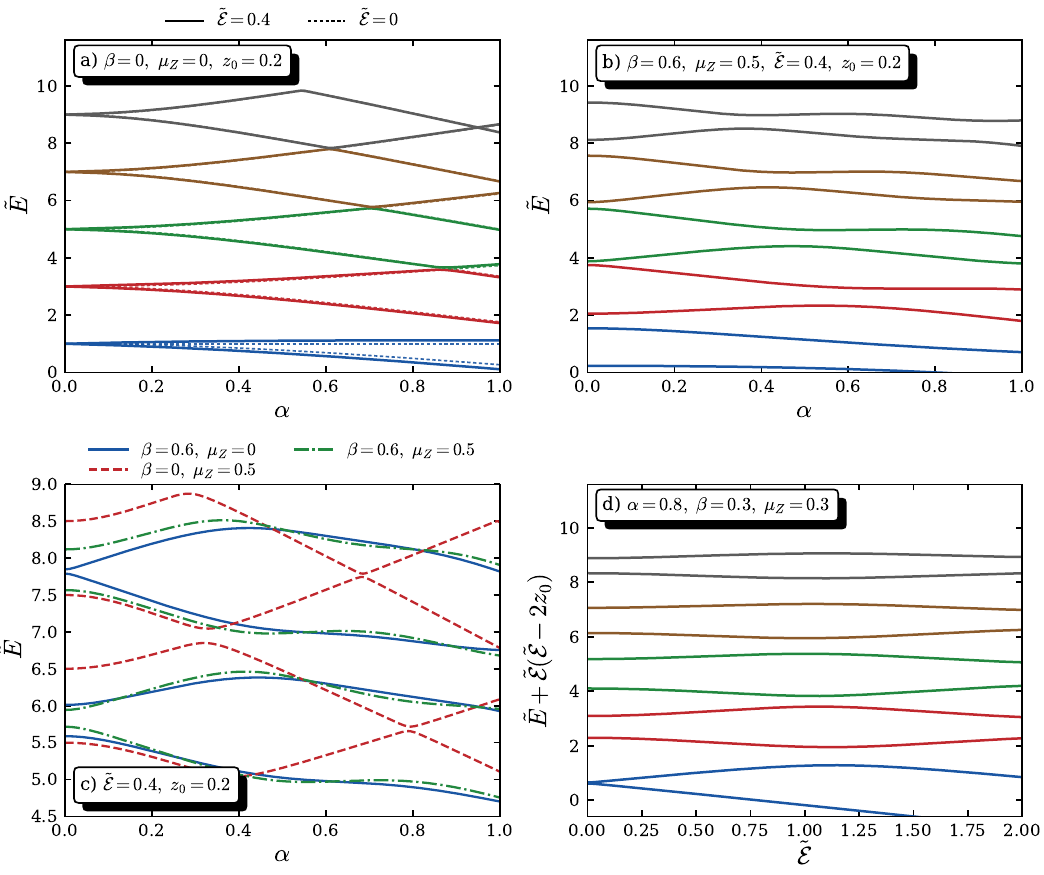}
    \caption{Spin-orbit coupled energy spectrum 
    in crossed fields. (a) Comparison of the 
    finite and zero electric field spectra for 
    the pure Rashba case. (b) Spectrum in the 
    presence of Dresselhaus and Zeeman terms. 
    (c) Comparison of the effects of the 
    Dresselhaus and Zeeman couplings near 
    the level anticrossings. (d) Dependence 
    of the spectrum on the electric field.}
    \label{fig:spectrum}
\end{figure*}

Having established these analytical limits, we
examine the full spectrum from 
Eqs.~\eqref{eq:recursion}. Fig.~\ref{fig:spectrum} 
shows the numerical results for $N=80$ Hermite 
functions, for which the twelve lowest levels are 
converged to better than $10^{-10}$. For the 
spectra shown as a function of $\alpha$, we choose 
$\tilde{\mathcal E}=0.4$ and $z_0=0.2$, which 
satisfy $\tilde{\mathcal E}^2=
2\tilde{\mathcal E}z_0$. With this choice, 
$\tilde E=\mathbb E_n+1$ and the classical drift 
terms $-m^*v_d^2/2+|e|\mathcal E x_c$ in 
Eq.~\eqref{eq:Enky} cancel.

Fig.~\ref{fig:spectrum}(a) compares, for 
$\beta=\mu_Z=0$, the finite electric field 
spectrum obtained by direct diagonalization 
of Eqs.~\eqref{eq:recursion} with the zero 
field Bychkov--Rashba spectrum of 
Eq.~\eqref{eq:ByRas}. The finite field results 
are numerically exact within the convergence 
accuracy of the Hermite basis. For weak Rashba 
coupling, the two spectra remain close, but the 
drift field $-m^*v_d(\alpha_R\sigma_x-
\beta_D\sigma_y)$ removes the simple zero-field 
structure. In particular, the isolated level 
$E_{\rm iso}$, which is independent of $\alpha$ 
at $\tilde{\mathcal E}=0$, is shifted upward, 
while the lower state of the first doublet is 
shifted downward. Since the Rashba levels are 
nearly spin degenerate at small $\alpha$, the 
drift field produces a first-order splitting 
of about $\pm\alpha\tilde{\mathcal E}$, or 
$\pm m^*v_d\alpha_R$ in physical units. When 
the Rashba splitting becomes larger than the 
drift energy $\alpha\tilde{\mathcal E}$, the 
field-induced shifts become predominantly 
quadratic.
A more pronounced effect appears near the 
crossings of the zero-field Bychkov--Rashba 
spectrum. For example, the lower state of 
the doublet $j+2$ crosses the upper state 
of the doublet $j$, with the first crossing 
occurring at $\alpha=0.888$ for $j=0$. Since 
the two states differ by two oscillator 
quanta, they are not coupled directly by 
the drift field. Instead, one state first 
couples to a state of a neighboring doublet, 
and this intermediate state then couples to 
the second crossing state. Since each 
coupling is proportional to 
$\tilde{\mathcal E}$, the resulting 
interaction between the two crossing states 
is proportional to $\tilde{\mathcal E}^2$. 
As a result, each crossing is replaced 
by a narrow anticrossing. These gaps are 
small on the overall energy scale of 
Fig.~\ref{fig:spectrum}(a), but their 
systematic field dependence provides a 
clear signature of the drift field. 
For $\tilde{\mathcal E}=0.4$, the three 
lowest anticrossing gaps are $0.080$, 
$0.052$ and $0.039$. Since 
$\tilde{\mathcal E}^2=0.16$, these values 
correspond to $\Delta_{\rm gap}/
\tilde{\mathcal E}^2\simeq0.50$, $0.33$ 
and $0.24$, respectively. We repeated 
the calculation for weaker electric 
fields down to $\tilde{\mathcal E}=0.05$ 
and found that these ratios remain nearly 
constant. This confirms that the anticrossing 
gaps scale quadratically with the electric 
field, $\Delta_{\rm gap}\propto
\tilde{\mathcal E}^2$.

The Dresselhaus coupling changes the 
spectrum more strongly. In contrast to the 
pure Rashba case, it directly mixes 
different Landau doublets already at zero 
electric field, so that the level crossings 
are replaced by much wider anticrossings. 
This behavior is clearly seen in 
Fig.~\ref{fig:spectrum}(b) for $\beta=0.6$ 
and $\mu_Z=0.5$, where the simple 
Bychkov--Rashba structure is strongly 
modified. The electric field then acts on 
these already mixed states through the 
drift field. This Landau level mixing 
produced by the simultaneous presence 
of Rashba and Dresselhaus couplings is 
consistent with the zero-field results 
of Refs.~\cite{Schliemann2003, Zarea2005} 
and remains present at finite electric 
field.

Fig.~\ref{fig:spectrum}(c) separates the 
effects of the Zeeman and Dresselhaus 
terms in the energy window $4.5<\tilde 
E<9$, where levels belonging to different 
doublets approach each other. 
For $\beta=0$ and $\mu_Z=0.5$, the Zeeman 
term changes mainly the positions of 
the avoided crossings, while the gaps 
themselves are still opened only by the 
drift field and remain small, with 
minimum separations between about 
$0.04$ and $0.23$. When the Dresselhaus 
coupling is included with $\beta=0.6$, 
it directly mixes the different doublets 
and the corresponding gaps increase 
strongly to values between about $0.5$ 
and $1.2$. Thus, the Dresselhaus 
coupling changes the anticrossing 
scale from small drift field induced 
gaps to gaps of order unity, while 
the Zeeman term mainly shifts their 
positions. The anticrossing width 
therefore provides a direct measure 
of the Dresselhaus coupling.

The dependence on the electric field 
is shown in Fig.~\ref{fig:spectrum}(d) 
for $\alpha=0.8$, $\beta=0.3$ and 
$\mu_Z=0.3$. At $\tilde{\mathcal E}=0$, 
the two lowest levels are nearly 
degenerate, with $\mathbb E_n+1=0.61$ 
and $0.65$. Since the drift field 
couples these two states directly, 
their energies separate linearly as 
the electric field is turned on. 
In contrast, the higher levels are 
not directly coupled within each 
doublet and therefore show mainly 
quadratic shifts, with the two 
members bending in opposite 
directions. As $\tilde{\mathcal E}$ 
increases, the drift field energy 
$\tilde{\mathcal E}\sqrt{\alpha^2+\beta^2}$ 
becomes comparable to the separation 
between neighboring levels. The mixing 
between these levels then becomes 
stronger and produces a sequence of 
avoided crossings. At still larger fields, 
the lowest level is governed mainly 
by the drift field and decreases 
approximately linearly with 
$\tilde{\mathcal E}$, approaching the 
asymptotic behavior $-\tilde{\mathcal E}
\sqrt{\alpha^2+\beta^2}$.

\begin{figure}
    \centering
    \includegraphics[width=0.8\textwidth]
    {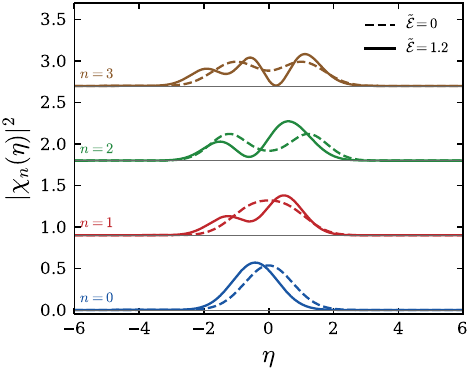}
    \caption{Spinor densities 
    $|\chi_n(\eta)|^2$ of the four lowest 
    states for $\tilde{\mathcal E}=0$ and 
    $1.2$, with $\alpha=0.8$, $\beta=0.3$ 
    and $\mu_Z=0.3$. The curves for different 
    $n$ are vertically shifted for clarity.}
    \label{fig:density}
\end{figure}

Finally, we examine the effect of the 
electric field on the spinor densities 
$|\chi_n(\eta)|^2=e^{-\eta^2}
\left(|\varphi_{n\uparrow}(\eta)|^2
+|\varphi_{n\downarrow}(\eta)|^2\right)$. 
The corresponding densities for the 
four lowest states are shown in 
Fig.~\ref{fig:density}. At 
$\tilde{\mathcal E}=0$, the densities 
are symmetric about $\eta=0$ because 
the spin components contain oscillator 
functions with definite parity. A 
finite electric field introduces the 
drift field, which breaks this 
symmetry and changes both the position 
and the shape of the densities. For 
$\tilde{\mathcal E}=1.2$, the centers 
of the four lowest states are shifted 
to $\langle\eta\rangle_n=-0.42$, 
$-0.06$, $0.07$ and $-0.14$ for 
$n=0,\ldots,3$. Thus, the electric 
field modifies not only the energy 
spectrum but also the spatial 
structure of the spinor states, which, 
as shown in Sec.~\ref{sec:3f}, is directly 
reflected in the electroabsorption 
line shapes.

\subsection{\label{sec:3b}Spin dynamics in a pure 
electric field}

We now consider the case $B=0$, where the Landau levels 
disappear, and the electric field drives the electron 
continuously along the $k_x$ direction according to
$\hbar\,dk_x/dt=-|e|\mathcal E$.
Since $k_y$ remains conserved, the changing $k_x$ also 
changes the Rashba-Dresselhaus spin-orbit field. The 
main question is therefore whether the spin follows 
this changing field or remains approximately fixed 
as $k_x$ varies.

For $B=0$, the Hamiltonian becomes
\begin{eqnarray}
H =
\frac{p^2}{2m^*}
+|e|\mathcal E x  +\hbar
\left[
(\alpha_R k_y+\beta_D k_x)\sigma_x
-(\alpha_R k_x+\beta_D k_y)\sigma_y
\right]. \ \ \
\label{eq:H_E}
\end{eqnarray}
Since the electric potential is linear in $x$, it is 
convenient to work in the $k_x$ representation, 
where $x\rightarrow i\partial/\partial k_x$. The 
Schr\"odinger equation then takes the form
\begin{equation}
i|e|\mathcal E
\frac{\partial\phi}{\partial k_x}
=
\left[
\varepsilon
-\frac{\hbar^2(k_x^2+k_y^2)}{2m^*}
-H_{\rm SO}
\right]\phi .
\label{eq:kx}
\end{equation}
To describe the spin part, we introduce
$g=\sqrt{\alpha_R^2+\beta_D^2}$ and choose the 
spin quantization axis along
$\mathbf n_1=(\beta_D,-\alpha_R)/g$.
In this basis the spin-orbit interaction 
becomes
\begin{equation}
H_{\rm SO}
=
\hbar
\left(
gk_x+\frac{2\alpha_R\beta_D}{g}k_y
\right)\tau_z
+
\hbar k_y
\frac{|\alpha_R^2-\beta_D^2|}{g}
\tau_x .
\label{eq:HSO_LZ}
\end{equation}
The first term changes continuously with 
$k_x$, while the second term couples the 
two spin states.
Since the kinetic energy acts equally on 
both spin components, it only produces a 
common phase and does not affect the spin 
dynamics. This phase can therefore be 
removed. The remaining spinor $\chi(k_x)$ 
obeys
\begin{equation}
i|e|\mathcal E
\frac{d\chi}{dk_x}
=
-
\left[
\Lambda(k_x)\tau_z
+\Gamma\tau_x
\right]\chi ,
\label{eq:LZ}
\end{equation}
where
$
\Lambda(k_x)
=
\hbar
\left(
gk_x+2\alpha_R\beta_Dk_y/g
\right),
\Gamma=
\hbar k_y
|\alpha_R^2-\beta_D^2|/g.$
Eq.~\eqref{eq:LZ} is exactly the Landau--Zener 
problem. Here the electric field continuously 
changes $k_x$, while $\Gamma$ determines the 
coupling between the two spin states.

The probability that the spin remains in the 
same $\tau_z$ state during the electric field 
driven passage through the crossing is
\begin{equation}
P_{\rm dia}
=
e^{-\pi\delta_{\rm LZ}},
\qquad
\delta_{\rm LZ}
=
\frac{
\hbar k_y^2
(\alpha_R^2-\beta_D^2)^2
}{
|e|\mathcal E
(\alpha_R^2+\beta_D^2)^{3/2}
}.
\label{eq:LZprob}
\end{equation}
The parameter $\delta_{\rm LZ}$ gives a 
direct criterion for the spin response 
to the electric field. For 
$\delta_{\rm LZ}\ll1$, which occurs for 
a strong electric field, small $k_y$, or 
$\alpha_R\simeq\beta_D$, the electric 
field drives $k_x$ so rapidly that the 
spin cannot follow the changing spin 
orbit field. The spin therefore remains 
approximately fixed and the dynamics is 
diabatic. In contrast, for 
$\delta_{\rm LZ}\gg1$, $k_x$ changes 
slowly enough that the spin continuously 
follows the local spin-orbit field, 
corresponding to the adiabatic regime.

This result shows that the electric field 
controls a crossover between frozen spin 
dynamics and adiabatic spin following. 
The crossover is governed by the combined 
effect of $\mathcal E$, $k_y$, and the 
relative strengths of the Rashba and 
Dresselhaus couplings.

It is also useful to express this result 
in the natural scales of the electric 
field problem. Introducing
$\ell=(\hbar^2/2m^*|e|\mathcal E)^{1/3}$ and
$\Theta=|e|\mathcal E\ell$, together with
$a=\hbar\alpha_R/(\ell\Theta)$ and
$b=\hbar\beta_D/(\ell\Theta)$, we obtain
\begin{equation}
\delta_{\rm LZ}
=
(k_y\ell)^2
\frac{(a^2-b^2)^2}
{(a^2+b^2)^{3/2}} .
\end{equation}
Since $\Theta\propto\mathcal E^{2/3}$, increasing 
the electric field drives the system toward the 
diabatic regime, while decreasing it leads to 
the adiabatic regime.

A particularly simple limit appears for 
$\alpha_R=\beta_D$. In this case $\Gamma=0$ 
for every $k_y$, so the two spin projections 
decouple exactly, and the spin does not rotate 
during the motion. The problem then reduces 
to two independent Airy equations. Explicitly, 
the spin-orbit term becomes $\sqrt2\hbar
\alpha(k_x+k_y)\tau$ with $\tau=(\sigma_x-
\sigma_y)/\sqrt2$, and the two projections 
$\tau=\pm1$ have the dispersions 
$\varepsilon_\tau(\mathbf k)=\hbar^2[(k_x
+\tau k_0)^2+(k_y+\tau k_0)^2]/2m^*
-2m^*\alpha^2$ with $k_0=\sqrt2m^*\alpha/
\hbar$. The displacement along $k_x$ is a 
gauge transformation $e^{-i\tau k_0x}$ that 
leaves the probability density unchanged, 
so that the eigenfunctions are the Airy 
functions of the spin-degenerate problem 
with the energy shifted by $-2m^*\alpha^2
+\hbar^2(k_y+\tau k_0)^2/2m^*$. This is 
the persistent spin helix point, where the 
spin-orbit field has a fixed direction and 
the electric field problem becomes exactly 
solvable.

Finally, a linear electric potential in an 
infinite crystal does not produce bound states. 
The spectrum therefore remains continuous, 
and the main result of this subsection is the 
electric field-controlled spin dynamics rather 
than a discrete energy spectrum.

\subsection{\label{sec:3c}Spin-orbit induced subgap 
absorption}

From this point onward, we focus on the optical 
response of the model.
For this purpose, we first consider the case 
in which both the electric and magnetic fields 
are absent, where the eigenfunctions are plane 
waves and $|\Psi_\nu(0)|^2=1/A$, with $A$ denoting 
the area of the crystal. Eq.~\eqref{eq:jcv} then 
reduces to
$
J_{cv}(\hbar\omega)
=
\frac{1}{A}
\sum_{\mathbf k,s}
\delta
\left(
E_{cv}^{s}(\mathbf k)-\hbar\omega
\right),
$
where $s=\pm$ labels the two spin-split branches, 
with $s=+$ denoting the upper branch and $s=-$ the 
lower branch, as illustrated in 
Fig.~\ref{fig:model}(c) for the pure Rashba case. 
Their dispersions are
\begin{equation}
E_{cv}^{s}(k,\theta)
=
E_g
+
\frac{\hbar^2k^2}{2m^*}
+
s\hbar k g(\theta),
\label{eq:dispersion_SOCs}
\end{equation}
where $
g(\theta)
=
\sqrt{
\alpha_R^2+\beta_D^2
+
2\alpha_R\beta_D\sin2\theta
},$ and $\theta$ is the polar angle of $\mathbf k$. 
It is important to emphasize here that, for $s=+$, 
the spin-orbit term increases the energy with $k$, 
so the minimum remains at $k=0$ and no additional 
minima or saddle points appear. The situation is 
different for the lower branch $s=-$, where the 
spin-orbit term lowers the energy and produces 
minima at finite $k$.
To locate these minima, it is sufficient to consider 
the dispersion for a fixed direction $\theta$,
$E_{cv}^{-}(k,\theta)=E_g+\hbar^2k^2/(2m^*)-\hbar 
k g(\theta)$.
Its minimum occurs at $k=m^*g(\theta)/\hbar$, 
and the corresponding energy measured from $E_g$ 
is $-m^*g^2(\theta)/2$. As can be seen, the angular 
dependence of the band minimum is therefore 
determined entirely by $g(\theta)$. This function 
reaches its largest value $\alpha_R+\beta_D$ at 
$\theta=\pi/4$ and $5\pi/4$. These directions 
therefore correspond to the deepest points of 
the lower branch, with
$k=m^*(\alpha_R+\beta_D)/\hbar$
and energy
$-m^*(\alpha_R+\beta_D)^2/2$.
In contrast, $g(\theta)$ reaches its smallest 
value $|\alpha_R-\beta_D|$ at $\theta=3\pi/4$ 
and $7\pi/4$. For each of these directions, 
the energy is still minimized with respect to 
$k$. However, since $g(\theta)$ is smallest 
there, the corresponding minimum energy 
$-m^*g^2(\theta)/2$ is higher than for slightly 
different values of $\theta$. Thus, the energy 
is a minimum along the radial direction $k$ but 
a maximum along the angular direction $\theta$. 
These points are therefore saddle points, with
$k=m^*|\alpha_R-\beta_D|/\hbar$
and energy
$-m^*(\alpha_R-\beta_D)^2/2$.
We also note that, for pure Rashba coupling, 
$\beta_D=0$, the function $g(\theta)=\alpha_R$ 
becomes independent of $\theta$. The minimum 
then has the same energy for every direction, 
so the isolated minima and saddle points merge 
into the continuous ring-shaped minimum at 
$-E_{\rm SO}$ shown in Fig.~\ref{fig:model}(c).

In light of these observations, we evaluate 
the JDOS from Eq.~\eqref{eq:dispersion_SOCs} 
by introducing $\Delta=\hbar\omega-E_g$. The 
sign of $\Delta$ determines which spin branches 
contribute to the JDOS. For $\Delta>0$, both 
branches contribute, and each branch has one 
radial solution for a given $\theta$. For 
$\Delta<0$, however, the upper branch does not 
contribute, while the lower branch has two 
radial solutions
$
k_\pm
=
\frac{m^*}{\hbar}
\left[
g(\theta)
\pm
\sqrt{
g^2(\theta)+\frac{2\Delta}{m^*}
}
\right],
$
provided that
$g^2(\theta)>-2\Delta/m^*$.
After the radial integration, the remaining 
angular integral gives
\begin{equation}
J_{cv}^{s}(\Delta>0)
=
\frac{m^*}{2\pi\hbar^2}
\left[
1
-
\frac{s}{2\pi}
\int_0^{2\pi}
\frac{
g(\theta)\,d\theta
}{
\sqrt{
g^2(\theta)+2\Delta/m^*
}
}
\right],
\label{eq:Delta>0_SOCs}
\end{equation}
while below the gap
\begin{equation}
J_{cv}^{+}(\Delta<0)=0,
\end{equation}
and
\begin{equation}
J_{cv}^{-}(\Delta<0)
=
\frac{m^*}{2\pi^2\hbar^2}
\int_{g^2>-2\Delta/m^*}
\frac{
g(\theta)\,d\theta
}{
\sqrt{
g^2(\theta)+2\Delta/m^*
}
}.
\label{eq:jdos_zero}
\end{equation}
From Eq.~\eqref{eq:Delta>0_SOCs}, it can be 
seen directly that for $\Delta>0$ the 
total JDOS becomes
$J_{cv}=J_{cv}^{+}+J_{cv}^{-}=m^*/
(\pi\hbar^2)$.
In other words, within the constant matrix 
element approximation, the Rashba and 
Dresselhaus couplings do not change the total 
optical absorption above $E_g$.
This sum rule holds for arbitrary values of 
the two couplings and is the optical 
counterpart of the well-known property of 
the density of states of a two-dimensional 
Rashba gas, which is constant above the 
band crossing point \cite{Winkler2014}. The 
couplings only redistribute the weight 
between the two branches, which approach 
$m^*/2\pi\hbar^2$ from opposite sides for 
$\Delta\rightarrow\infty$. 
We therefore continue with the case $\Delta<0$. 
Since only the lower branch $s=-$ contributes 
in this region, the lowest possible transition 
energy is determined by the absolute minimum 
of this branch, which is
$E_g-m^*(\alpha_R+\beta_D)^2/2$.
The relevant sub-gap region can thus be written 
as
$
E_g-m^*(\alpha_R+\beta_D)^2/2
<
\hbar\omega
<
E_g.
$

We first consider the pure Rashba case, 
$\beta_D=0$. In this limit, Eq.~\eqref{eq:jdos_zero} 
can be evaluated analytically and gives
\begin{equation}
J_{cv}^{-}(\Delta)
=
\frac{m^*}{\pi\hbar^2}
\left(
1+\frac{\Delta}{E_{\rm SO}}
\right)^{-1/2}.
\label{eq:jdos_rashba}
\end{equation}
The ring-shaped minimum of the lower Rashba 
branch therefore produces an inverse square 
root van Hove singularity at $\hbar\omega=
E_g-E_{\rm SO}$, which appears in the optical 
response as a divergent subgap absorption edge. 
The situation changes qualitatively when the 
Dresselhaus coupling is added. The continuous 
Rashba ring minimum is then broken into two 
isolated minima and two saddle points. As a 
result, the inverse square root singularity 
is replaced by two separate features. A finite 
step appears at
$
\Delta
=
-m^*(\alpha_R+\beta_D)^2/2,
$
which corresponds to the absolute minima of the 
lower branch, while a logarithmic singularity 
appears at
$
\Delta
=
-m^*(\alpha_R-\beta_D)^2/2,
$
which corresponds to the saddle points.

\begin{figure}
    \centering
    \includegraphics[width=0.48\textwidth]
    {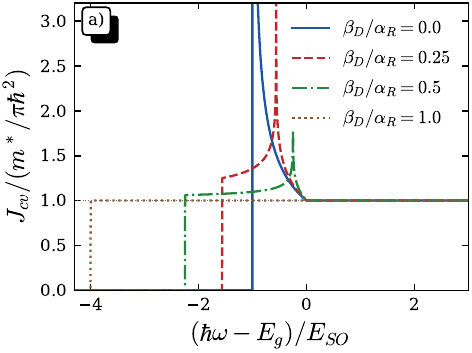}
    \hfill
    \includegraphics[width=0.48\textwidth]
    {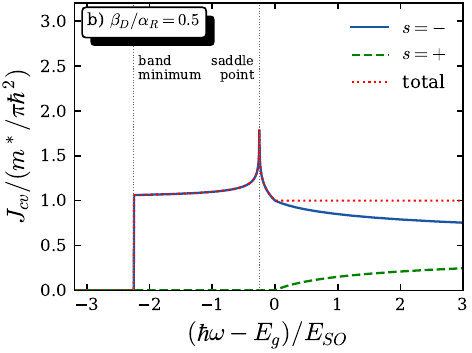}
    \caption{Joint density of states at zero 
    electric and magnetic fields. (a) Total 
    JDOS for several values of 
    $\beta_D/\alpha_R$. (b) Contributions 
    of the $s=\pm$ branches and their 
    sum for $\beta_D/\alpha_R=0.5$.}
    \label{fig:JODS_SOCs}
\end{figure}

These results are shown in 
Fig.~\ref{fig:JODS_SOCs}. 
Fig.~\ref{fig:JODS_SOCs}(a) presents the total 
JDOS for several values of $\beta_D/\alpha_R$, 
while panel (b) shows the $ s=+$ separately 
and $s=-$ contributions and their sum for 
$\beta_D/\alpha_R=0.5$.
All curves in Fig.~\ref{fig:JODS_SOCs}(a) 
become identical above the gap, confirming 
that the total JDOS is independent of the 
spin-orbit coupling in this region. For pure 
Rashba coupling, the lower absorption edge 
contains the inverse square root singularity 
associated with the ring minimum, shown by 
the solid blue curve in 
Fig.~\ref{fig:JODS_SOCs}(a). For 
$\beta_D/\alpha_R=0.25$ and $0.5$, shown 
by the dashed red and dash-dotted green 
curves, respectively, this singularity is 
replaced by a finite step with heights 
$1.25$ and $1.06$ in units of $m^*/\pi\hbar^2$, 
followed by a logarithmic peak at the saddle 
point energy. It is important to note that the 
height of this step can also be obtained 
analytically by expanding 
Eq.~\eqref{eq:jdos_zero} near the band minimum, 
which gives
\begin{equation}
J_{cv}
\left(
\Delta
\rightarrow
-\frac{m^*(\alpha_R+\beta_D)^2}{2}
+0
\right)
=
\frac{m^*}{2\pi\hbar^2}
\frac{\alpha_R+\beta_D}
{\sqrt{\alpha_R\beta_D}}.
\end{equation}
This expression gives the values $1.25$ and 
$1.06$ for $\beta_D/\alpha_R=0.25$ and $0.5$, 
respectively, in agreement with 
Fig.~\ref{fig:JODS_SOCs}(a). At 
$\beta_D=\alpha_R$, the saddle point structure 
disappears and the JDOS reduces to the spin 
degenerate step
$J_{cv}=(m^*/\pi\hbar^2)
\Theta_H(\Delta+4E_{\rm SO})$,
shown by the dotted brown curve in 
Fig.~\ref{fig:JODS_SOCs}(a). This value also 
follows directly from the expression above, 
which gives the spin-degenerate step height 
$m^*/\pi\hbar^2$ at $\beta_D=\alpha_R$. In 
the opposite limit $\beta_D\rightarrow0$, 
the step height diverges, continuously 
recovering the inverse square root 
singularity of the pure Rashba case in 
Eq.~\eqref{eq:jdos_rashba}.

The subgap structure therefore contains 
direct information about both spin-orbit 
couplings. The position of the lower step 
determines $(\alpha_R+\beta_D)^2$, while 
the logarithmic peak determines 
$(\alpha_R-\beta_D)^2$. Measuring both 
energies allows $\alpha_R$ and $\beta_D$, 
and therefore their ratio, to be extracted 
separately from a single optical absorption 
spectrum. This provides a direct optical 
method for determining the two spin-orbit 
couplings without applying any external 
electric or magnetic field.

\subsection{\label{sec:3d}Spin-orbit coupled 
Franz--Keldysh effect}

We now examine how the optical response associated with 
the SOCs discussed in the previous subsection changes in 
the presence of an electric field. As emphasized in 
Sec.~\ref{sec:3b}, the motion along the $x$ direction 
is unbounded and the states form a continuum, while 
$k_y$ remains conserved. Therefore, the JDOS can be 
written as
\begin{equation}
J_{cv}(\Delta)
=
\int
\frac{dk_y}{2\pi}
\rho_{k_y}(0,\Delta),
\label{eq:jdos_E}
\end{equation}
where
$
\rho_{k_y}(x,\varepsilon)
=
-\frac{1}{\pi}
{\rm Im}\,
{\rm Tr}\,
G_{k_y}(x,x,\varepsilon+i0)$
is the spin-traced local density of states for a 
fixed $k_y$ and
$G_{k_y}$ is the retarded Green's function of the 
one-dimensional problem at fixed $k_y$, with 
the energy $\varepsilon$ measured from $E_g$. 
We first emphasize that, in the absence of spin 
orbit coupling, the motion along the $x$ direction 
reduces to the usual problem of a particle in a 
linear potential, and the corresponding solutions 
are Airy functions. The local density of states at 
the origin is then
$
\rho_{k_y}(0,\varepsilon)
=
\frac{2}{\Theta\ell}
{\rm Ai}^2
\left(
\frac{
\hbar^2k_y^2/(2m^*)-\varepsilon
}{
\Theta
}
\right),$ where $\ell$ and $\Theta$ are the length 
and energy scales introduced in Sec.~\ref{sec:3b} 
and the factor $2$ accounts for the two spin states.
After integration over $k_y$, the JDOS is
\begin{equation}
J_{cv}(\Delta)
=
\frac{2m^*}{\pi\hbar^2}
F_{\rm 2D}
\left(
\frac{\Delta}{\Theta}
\right),
\label{eq:FK2D}
\end{equation}
where
$
F_{\rm 2D}(\eta)
=
\int_{-\eta}^{\infty}
\frac{{\rm Ai}^2(t)\,dt}
{\sqrt{t+\eta}}
=
2
\int_0^\infty
{\rm Ai}^2(u^2-\eta)\,du .
$
For large positive $\eta$, $F_{\rm 2D}
\rightarrow1/2$, and therefore the JDOS 
approaches the field-free value $m^*/\pi\hbar^2$.
Eq.~\eqref{eq:FK2D} describes the two dimensional 
Franz--Keldysh effect. The electric field allows 
finite absorption below $E_g$, where the 
absorption decreases rapidly as the photon 
energy moves further below the gap. Above $E_g$, 
the absorption oscillates around the field-free 
value, with the oscillation amplitude gradually 
decreasing at higher energies. Both effects 
occur over the characteristic energy scale 
$\Theta$.

For finite SOCs, we first consider the persistent 
spin helix point $\alpha_R=\beta_D\equiv\alpha$. 
As shown in Sec.~\ref{sec:3b}, the two spin 
projections decouple exactly in this case and 
the spin-orbit coupling produces only an energy 
shift. Therefore,
\begin{equation}
J_{cv}(\Delta)
=
\frac{2m^*}{\pi\hbar^2}
F_{\rm 2D}
\left(
\frac{
\Delta+2m^*\alpha^2
}{
\Theta
}
\right).
\label{eq:FK_PSH}
\end{equation}
Both spin projections give the same contribution 
and the entire Franz--Keldysh spectrum is shifted 
by $2m^*\alpha^2$. This is consistent with the 
field free result of Sec.~\ref{sec:3c}, where 
the absorption edge is shifted to $\Delta=
-2m^*\alpha^2$.

\begin{figure*}
    \centering
    \includegraphics[width=1\textwidth]
    {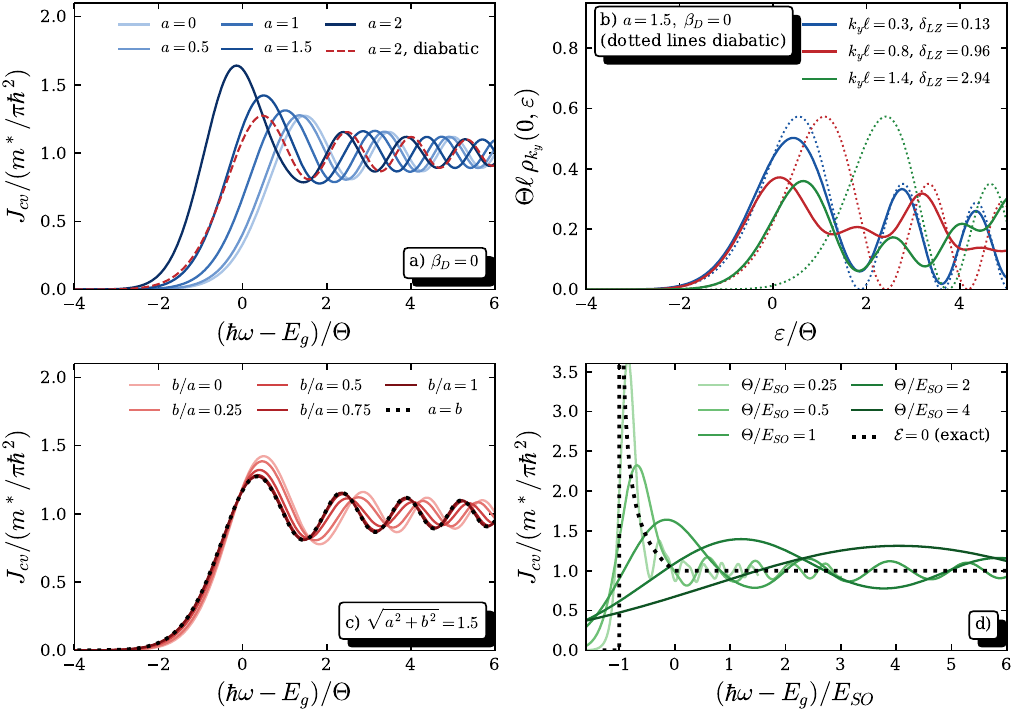}
    \caption{Franz--Keldysh electroabsorption 
    with spin-orbit coupling at $B=0$. (a) 
    JDOS for pure Rashba coupling. (b) Local 
    density of states $\Theta\ell\rho_{k_y}
    (0,\varepsilon)$ for different $k_y\ell$. 
    (c) JDOS for different Rashba and Dresselhaus 
    coupling ratios at fixed $\sqrt{a^2+b^2}=1.5$. 
    (d) JDOS for a fixed material and different 
    electric field strengths.}
    \label{fig:FK}
\end{figure*}

For general $\alpha_R\neq\beta_D$ the situation
is different. The Landau--Zener coupling derived
in Sec.~\ref{sec:3b} mixes the two spin states 
and a closed analytical expression is no longer 
available. We therefore calculate 
Eq.~\eqref{eq:jdos_E} numerically using the 
real-space Green function. The corresponding 
results are shown in Fig.~\ref{fig:FK}. 
Fig.~\ref{fig:FK}(a) shows the combined 
effect of the Rashba spin-orbit coupling and 
the electric field for $\beta_D=0$, where the 
strength of the Rashba coupling is described 
by $a=\hbar\alpha_R/(\ell\Theta)$.
As $a$ increases, the absorption edge moves 
downward by
$E_{\rm SO}=a^2\Theta/4$, in agreement with 
the field-free threshold found in 
Sec.~\ref{sec:3c}. At the same time, the first 
Franz--Keldysh maximum increases from
$1.27\,m^*/\pi\hbar^2$ for $a=0$
to $1.42\,m^*/\pi\hbar^2$ for $a=1.5$
and $1.64\,m^*/\pi\hbar^2$ for $a=2$.
The increase of this peak reflects the inverse 
square-root van Hove singularity of the 
field-free Rashba system.

The role of the Landau--Zener spin dynamics 
discussed in Sec.~\ref{sec:3b} is shown in 
Fig.~\ref{fig:FK}(b). The solid curves give the 
exact local density of states, while the dotted 
curves correspond to the diabatic approximation, 
where the spin remains frozen as $k_x$ changes. 
For $k_y\ell=0.3$, where $\delta_{\rm LZ}=0.13$, 
the two results are close, as expected in the 
diabatic regime. At $k_y\ell=0.8$, where 
$\delta_{\rm LZ}=0.96$, visible differences 
appear, indicating the crossover between the 
diabatic and adiabatic regimes. For $k_y\ell=1.4$, 
where $\delta_{\rm LZ}=2.94$, the exact result 
differs strongly from the diabatic one, with the 
absorption threshold shifted downward by about 
$1.5\Theta$ and a different oscillation structure. 
This shows that the Landau--Zener spin evolution 
becomes important for $a\gtrsim1$, corresponding 
approximately to $\Theta\lesssim4E_{\rm SO}$, 
and directly modifies the electroabsorption 
spectrum.

The effect of the Dresselhaus coupling is shown 
in Fig.~\ref{fig:FK}(c). Here the total spin 
orbit strength
$\sqrt{a^2+b^2}=1.5$
is kept fixed, where
$b=\hbar\beta_D/(\ell\Theta)$,
while the ratio of the Rashba and Dresselhaus 
couplings is varied. The changes in the spectrum 
are more moderate than in the pure Rashba case. 
In particular, the first maximum decreases from
$1.42\,m^*/\pi\hbar^2$
for $b=0$
to
$1.27\,m^*/\pi\hbar^2$
at $a=b$.
This moderate change reflects two competing 
effects. Increasing the Dresselhaus coupling 
lowers the band minimum, while at the same 
time it reduces the Landau--Zener spin mixing. 
At $a=b$, this mixing vanishes, and the spectrum 
reduces to the exact shifted Franz--Keldysh 
result of Eq.~\eqref{eq:FK_PSH}.

Fig.~\ref{fig:FK}(d) shows the field dependence 
for a fixed material, where the effective mass 
$m^*$ and the Rashba coupling $\alpha_R$, and 
therefore $E_{\rm SO}=m^*\alpha_R^2/2$, are kept 
fixed while the electric field is varied. 
Since $\Theta\propto\mathcal E^{2/3}$, changing 
$\mathcal E$ changes the ratio $\Theta/E_{\rm SO}$ 
and continuously connects the strong and weak field 
regimes discussed in Sec.~\ref{sec:3b}. For 
$\Theta\gg E_{\rm SO}$, the spectrum is close 
to the ordinary Franz--Keldysh result shifted 
by $E_{\rm SO}$. As the electric field decreases, 
the spin dynamics becomes increasingly adiabatic 
and the spectrum approaches the field-free Rashba 
result of Sec.~\ref{sec:3c}. In this limit, the 
finite electric field replaces the inverse square 
root singularity at $\Delta=-E_{\rm SO}$ by a 
finite peak whose width is of order $\Theta$. 
As $\Theta$ decreases, this peak becomes higher 
and moves toward the field-free absorption edge 
$\hbar\omega=E_g-E_{\rm SO}$. For $\Theta\lesssim 
E_{\rm SO}/4$, its height and position are well 
described by
\begin{equation}
J_{\rm max}
\simeq
1.83
\left(
\frac{E_{\rm SO}}{\Theta}
\right)^{1/2}
\frac{m^*}{\pi\hbar^2}
\propto
\mathcal E^{-1/3},
\label{eq:Jmax}
\end{equation}
and
\begin{equation}
\hbar\omega_{\rm max}
\simeq
E_g-E_{\rm SO}+0.6\Theta .
\label{eq:Emax}
\end{equation}
Thus, decreasing the electric field causes the peak 
to approach the Rashba edge while its height increases 
as $\mathcal E^{-1/3}$. The field dependence of both 
quantities therefore provides a direct optical measure 
of $E_{\rm SO}$. The corresponding electroabsorption 
$\Delta\mathcal A=\mathcal A(\mathcal E)-\mathcal A(0)$ 
is negative in an energy interval of order $\Theta$ 
above $E_g-E_{\rm SO}$ and positive below it, with an 
amplitude that also grows as $\mathcal E^{-1/3}$. 
This behavior is qualitatively different from the 
ordinary two-dimensional Franz--Keldysh effect, where 
the oscillation amplitude is nearly field independent, 
and from the three-dimensional case, where it scales 
as $\mathcal E^{1/3}$. The opposite $\mathcal E^{-1/3}$ 
dependence found here originates from the inverse square 
root van Hove singularity produced by the ring-shaped 
minimum of the lower Rashba branch.

\subsection{\label{sec:3e}Spin-orbit coupled 
magnetoabsorption}

\begin{figure}
    \centering
    \includegraphics[width=0.8\textwidth]
    {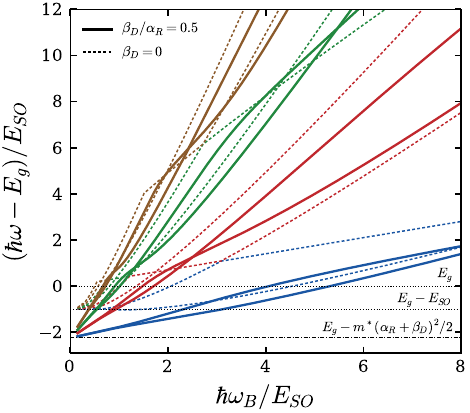}
    \caption{Positions $\hbar\omega=E_g+E_n$ 
    of the eight lowest magnetoabsorption 
    lines as a function of the magnetic 
    field for a fixed material, in units 
    of $E_{\rm SO}$, with $\mu_Z=0.3$.}
    \label{fig:magneto}
\end{figure}

We next consider the optical response in the presence of 
Rashba and Dresselhaus SOCs and a magnetic field, with 
$\mathcal E=0$. In this case, the states form 
spin-orbit-mixed Landau levels, and summing the degenerate 
$k_y$ states with the normalized spinor densities of 
Sec.~\ref{sec:3a} gives
\begin{equation}
J_{cv}(\hbar\omega)
=
\frac{1}{2\pi a_B^2}
\sum_n
\delta
\left(
\hbar\omega-E_g-E_n
\right),
\label{eq:JDOS_B}
\end{equation}
where $E_n=(\hbar\omega_B/2)(\mathbb E_n+1)$.
Eq.~\eqref{eq:JDOS_B} shows that each Landau level 
produces an absorption line at $\hbar\omega=E_g+E_n$, 
with the same total weight $1/(2\pi a_B^2)$, the 
Landau degeneracy per unit area. An important 
result is that this weight does not depend on how 
strongly the Rashba and Dresselhaus couplings mix 
the spin components. The spin-orbit coupling changes 
the energies $E_n$ and therefore shifts, splits, and 
produces anticrossings between the absorption lines, 
but it does not change their total weights. This is 
the magnetic field counterpart of the JDOS sum rule 
obtained in Sec.~\ref{sec:3c} and is exact within 
the constant matrix element approximation. The 
evolution of these lines with the magnetic field 
is shown in Fig.~\ref{fig:magneto}, which gives 
their positions for a fixed material as a function 
of $\hbar\omega_B\propto B$, with pure Rashba 
coupling (dashed lines) and with $\beta_D/\alpha_R
=0.5$ (solid lines). At weak magnetic fields, the 
absorption lines originate from the field-free 
edges obtained in Sec.~\ref{sec:3c}. These edges 
are $E_g-E_{\rm SO}$ for pure Rashba coupling and 
$E_g-m^*(\alpha_R+\beta_D)^2/2$ when the 
Dresselhaus coupling is also present. In the pure 
Rashba case, the lowest absorption line remains 
close to $E_g-E_{\rm SO}$ up to $\hbar\omega_B
\simeq2E_{\rm SO}$ because the lower branch of 
the $j$th Rashba doublet reaches its minimum 
near $-E_{\rm SO}$ when $\hbar\omega_B\simeq 
E_{\rm SO}/(j+1)$. In the pure Rashba case 
the splitting of a doublet grows as $\sqrt{B}$ 
at weak fields and becomes linear in $B$ at 
strong fields, where it approaches $\hbar\omega_B
-g\mu_BB$. The Dresselhaus coupling converts the 
crossings between different Rashba doublets into 
anticrossings, seen as a repulsion between the 
corresponding absorption lines. It also shifts 
the weak field lines to lower energies because 
the band minimum becomes deeper.

\subsection{\label{sec:3f}Spin-orbit coupled 
electroabsorption in crossed fields}

We finally consider the combined effect of Rashba and 
Dresselhaus SOCs and crossed electric and magnetic 
fields on the optical absorption. In crossed fields, 
the electric field removes the $k_y$ degeneracy of 
the Landau levels and makes the absorption depend 
directly on the spinor density. The JDOS becomes
\begin{equation}
J_{cv}(\hbar\omega)
=
\frac{1}{2\pi a_B^2}
\frac{1}{|e|\mathcal E a_B}
\sum_n
|\chi_n(\eta_n^*)|^2,
\label{eq:JDOS_crossed}
\end{equation}
where
$
\eta_n^*
=
\tilde{\mathcal E}
+
0.5\left(
\mathbb E_n+1-\tilde{\mathcal E}^2-\tilde{\Delta}
\right)/\tilde{\mathcal E}
$
and
$
\tilde{\Delta}
=
(\hbar\omega-E_g)/(\hbar\omega_B/2)
$.
Thus, the photon energy selects a position 
$\eta_n^*$ for each Landau level, and the absorption 
strength is determined by $|\chi_n(\eta_n^*)|^2$. 
The electric field therefore changes the shape and 
position of each absorption line while its 
integrated weight remains unchanged.

\begin{figure*}
    \centering
    \includegraphics[width=1\textwidth]
    {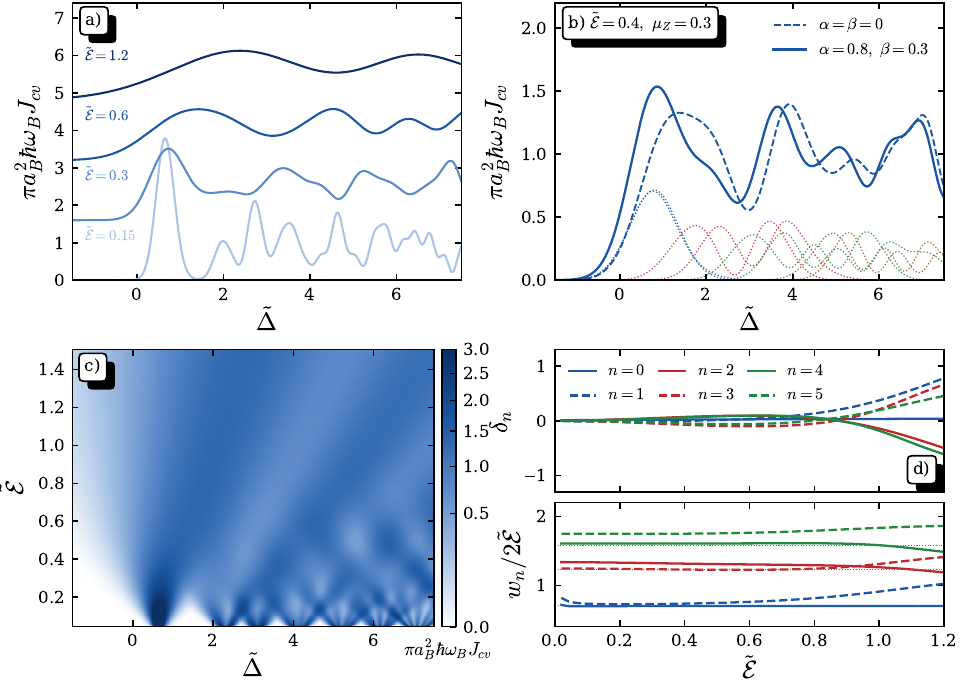}
    \caption{Electroabsorption in crossed electric 
    and magnetic fields for $\alpha=0.8$, $\beta=0.3$, 
    and $\mu_Z=0.3$. (a) JDOS for increasing 
    $\tilde{\mathcal E}$. (b) Comparison between the 
    cases with and without spin-orbit coupling at 
    $\tilde{\mathcal E}=0.4$. (c) JDOS as a function 
    of $\tilde{\Delta}$ and $\tilde{\mathcal E}$. 
    (d) Centroid shifts $\delta_n$ and reduced widths 
    $w_n/(2\tilde{\mathcal E})$ of the six lowest 
    absorption lines.}
    \label{fig:crossed}
\end{figure*}

Fig.~\ref{fig:crossed}(a) shows this evolution for 
$\alpha=0.8$, $\beta=0.3$, and $\mu_Z=0.3$. At 
$\tilde{\mathcal E}=0.15$ the Landau lines are 
narrow and well separated. Since the line profile 
follows the oscillator density $\phi_j^2$, a 
state with oscillator index $j$ produces $j+1$ 
maxima. The two lowest levels, which are nearly 
degenerate at $\tilde{\mathcal E}=0$ as shown in 
Fig.~\ref{fig:spectrum}(d), strongly overlap and 
appear as a single absorption line with 
approximately twice the spectral weight.
Increasing $\tilde{\mathcal E}$ broadens the 
absorption lines until the neighboring Landau-level 
contributions are no longer resolved separately, 
and at $\tilde{\mathcal E}=1.2$ the spectrum 
becomes Franz--Keldysh-like.

Fig.~\ref{fig:crossed}(b) compares the total JDOS 
with and without SOC at $\tilde{\mathcal E}=0.4$. 
The dashed curve shows the total spectrum without 
spin-orbit coupling; the solid curve shows the total spectrum with 
spin-orbit coupling, and the dotted curves represent the individual 
spin-orbit coupling mixed Landau-level contributions. Without spin-orbit coupling, 
each Landau line follows the oscillator density 
$\phi_j^2$, with centroid
$\langle\tilde{\Delta}\rangle=2j+1\pm\mu_Z
+\tilde{\mathcal E}^2$
and width
$|e|\mathcal E a_B\sqrt{j+1/2}$.
When spin-orbit coupling is present, different oscillator states are 
mixed, which shifts the individual absorption lines 
and modifies their widths and line profiles. As a 
result, the first maximum of the total spectrum 
moves from $\tilde{\Delta}=1.4$ without SOC to 
$\tilde{\Delta}=0.87$ with SOC.

Fig.~\ref{fig:crossed}(c) shows the JDOS as a 
two dimensional map of photon energy $\tilde{\Delta}$ 
and electric field $\tilde{\mathcal E}$, where darker 
regions correspond to stronger absorption. At small 
$\tilde{\mathcal E}$, the spectrum consists of narrow 
structures associated with individual Landau levels. 
As the electric field increases, these structures 
broaden and spread in energy until neighboring 
Landau-level contributions can no longer be 
distinguished separately. At larger $\tilde{\mathcal E}$, 
the electric-field broadening becomes comparable to 
or larger than the Landau-level spacing, so the 
individual magnetic absorption lines can no longer 
be resolved. The spectrum then approaches the $B=0$ 
Franz--Keldysh response. The relevant energy scale is
$\Theta=(\tilde{\mathcal E}^2/2)^{1/3}\hbar\omega_B$,
so this crossover occurs when $\Theta$ becomes larger 
than the Landau-level spacing.

Finally, Fig.~\ref{fig:crossed}(d) quantifies how SOC 
changes the individual absorption lines as the electric 
field increases. The upper panel shows the additional 
shift $\delta_n$ of the center of each absorption line, 
defined as
\begin{equation}
\delta_n
=
\langle\tilde{\Delta}\rangle_n
-
\left[\mathbb E_n(0)+1\right]
-
\tilde{\mathcal E}^2 .
\end{equation}
Here $\mathbb E_n(0)+1$ is the line position at zero 
electric field and $\tilde{\mathcal E}^2$ is the usual 
classical drift shift. Thus, $\delta_n>0$ corresponds 
to an additional shift to higher energy, while 
$\delta_n<0$ corresponds to a shift to lower energy. 
At weak electric fields, the two members of a 
spin-orbit mixed doublet are shifted in opposite directions. 
For example, at $\tilde{\mathcal E}=0.3$ we obtain 
$\delta_2=0.047$ and $\delta_3=-0.049$, showing that 
one line moves to higher energy while its partner 
moves to lower energy. For nondegenerate levels, 
these additional shifts vary as $\tilde{\mathcal E}^2$, 
which means that they arise as a second-order response 
to the electric field. The nearly degenerate lowest 
pair is different because the electric field mixes 
these two states already in first order, so their 
shifts start linearly in $\tilde{\mathcal E}$. This 
opposite motion of the two partners is a direct 
signature of the spin-orbit coupling induced drift field.
The lower panel shows the reduced line widths 
$w_n/(2\tilde{\mathcal E})$. The dotted horizontal 
lines give the corresponding widths without SOC, 
so the deviation of the colored curves from these 
lines directly measures the effect of spin mixing. 
SOC changes some of the widths by about $10\%$. 
At $\tilde{\mathcal E}\gtrsim0.8$, the field induced 
anticrossings discussed in Sec.~\ref{sec:3a} exchange 
the character of the levels, which is reflected in 
the crossing and rapid change of the width curves.

\section{\label{sec:Conclusions}Conclusions}

In this work, we have developed a unified description of the
interband optical response of a two-dimensional semiconductor
with Rashba and Dresselhaus spin-orbit couplings in electric
and magnetic fields. In crossed fields, the in-plane electric
field produces an additional spin-orbit drift field. This term
changes the spin-orbit-mixed Landau levels, shifts nearly
degenerate states, and produces field-induced anticrossings.
At zero magnetic field, the same electric field problem can
be mapped onto a Landau--Zener problem in momentum space.
This allows the strong-field diabatic and weak-field adiabatic
regimes to be clearly identified, together with the special
persistent spin helix limit.

The optical response shows several characteristic signatures
of the spin-orbit interaction. At zero external fields, the
total joint density of states above the band gap remains
unchanged, while new structures appear below the gap. Pure
Rashba coupling produces an inverse square root singularity,
whereas the addition of Dresselhaus coupling changes this
structure into a step and a logarithmic peak. In an electric
field, the Franz--Keldysh effect removes the Rashba singularity
and replaces it by a finite absorption peak. In the weak field
regime, the height of this peak varies as
$\mathcal E^{-1/3}$ and its position approaches
$E_g-E_{\rm SO}$ as the field is reduced. These dependences
provide a direct optical measure of the spin-orbit energy.
In a perpendicular magnetic field, the spin-orbit interaction
shifts and splits the magnetoabsorption lines while leaving
their total weights unchanged. When both fields are present,
the electric field broadens these lines and gradually drives
the spectrum toward the Franz--Keldysh regime. At the same
time, the two members of a spin-orbit-mixed Landau doublet
can acquire opposite shifts and different widths, providing
additional optical signatures of the underlying 
spin-orbit states.

It is also useful to connect these results with experimentally
accessible parameters. For a representative giant-Rashba
two dimensional semiconductor with $\mu=0.1 m_0$ and
$\hbar\alpha_R=0.5$~eV\,\AA, the corresponding spin orbit
energy is $E_{\rm SO}=1.64$~meV. At $B=5$~T, the 
spin-orbit-mixed Landau levels are already well separated, while electric
fields of a few kV/cm give a Franz--Keldysh energy scale
comparable to $E_{\rm SO}$. For these parameters, fields
between about $0.5$ and $3$~kV/cm are sufficient to broaden
the magnetoabsorption lines and drive the spectrum from
separated Landau structures toward a Franz--Keldysh-like
response. At zero magnetic field, fields of $1$--$4$~kV/cm
replace the field-free Rashba singularity with a measurable
subgap absorption peak whose position changes with the
electric field. These energy and field scales are relevant
for giant-Rashba materials such as polar semiconductors with
heavy elements~\cite{Ishizaka2011}. In conventional III--V
heterostructures, where the Rashba energy is much smaller,
the corresponding electroabsorption signatures are expected
to be weaker~\cite{Nitta1997,Vurgaftman2001}.

The predicted subgap absorption, its electric field dependence,
and the spin-dependent shifts and widths of the crossed field
absorption lines provide direct signatures that can be tested
by electroabsorption and magneto-optical spectroscopy. This
makes epitaxially grown III--V quantum wells a natural platform
for experimental verification, where gate electrodes can control
the in-plane electric field and optical measurements can follow
the resulting changes in the absorption spectrum. Such structures
can also be integrated with semiconductor laser and photodetector
technologies, opening a possible route toward electrically tunable
spin-dependent optical devices.


\ack{The authors declare that they have no competing financial interests or personal relationships that could have appeared to influence the work reported in this paper. BT is supported by the Turkish Academy of Sciences (TUBA) under Grant No. AD-2026.}
\roles{B.S.: Conceptualization, Methodology, Formal analysis, Writing -- original draft. S.T.: Formal analysis, Validation, Writing -- review and editing. B.T.: Conceptualization, Supervision, Writing -- review and editing.}
\data{The data that support the findings of this study are 
available from the authors upon reasonable
request.}

\bibliographystyle{iopart-num}
\bibliography{paper}

\end{document}